# Origin of non-Fermi liquid behavior in heavy fermion systems: A conceptual view

Swapnil Patil*

*Department of Condensed Matter Physics and Materials Science, Tata Institute of Fundamental Research, Homi Bhabha Road, Colaba, Mumbai 400005, India*

*Email: swapnil.patil@tifr.res.in

**Abstract**

We critically examine the non-Fermi liquid (NFL) behavior observed in heavy fermion systems located close to a magnetic instability and suggest a conceptual advance in physics in order to explain its origin. We argue that the treatment of electronic states responsible for magnetism near the Quantum Critical Point (QCP), should not be accomplished within the quantum mechanical formalism; instead they should be treated semi-classically. The observed NFL behavior can be explained within such a scenario. As a sequel we attempt to discuss its consequences for the explanation of high-$T_C$ superconductivity observed in Cuprates.

1.  **Introduction**

NFL behavior is observed in heavy fermion systems when the system is tuned towards a magnetic instability known as the QCP [1, 2]. Far away from QCP one observes normal Fermi liquid behavior as expected for metals by Landau [3]. Landau's Fermi liquid theory is remarkably successful in explaining the low temperature behavior of paramagnetic heavy fermion systems despite the strength of electronic correlations being as large as 6-7 eV when described within the Anderson model [4]. A big question is: What causes the breakdown of Landau's Fermi liquid theory close to the QCP while it remains valid far away from it. This has been a long standing problem within the scientific community and extensive experimental and theoretical investigations have been performed in order to shed light on it. Many theoretical progresses have been made in order to explain NFL behavior e.g. models based on multichannel Kondo effect, models based on QCP, models based on disorder etc. [1, 2]. Despite the intense study, a consensus over the origin of NFL behavior has not been reached yet.

In this paper, we propose a conceptual advance in physics over the treatment of electronic states close to the QCP. Our picture is based on certain overlooked issues in quantum mechanics which led to improper theoretical treatment of 'magnetic' states close to the QCP. These issues, when considered, readily explain the observed NFL behavior.

## 2. Results and Discussion

In order to illustrate our point, we take recourse to the single ion Kondo problem [5]. Let us assume a two electron system: a 'magnetic' $4f$ electron and an itinerant valence electron. The ground state for the single ion Kondo problem is the non-magnetic Kondo singlet state represented by $|\uparrow\rangle_f|\downarrow\rangle_v - |\downarrow\rangle_f|\uparrow\rangle_v$. This quantum mechanical state describes indistinguishable particles as is evident from its exchange symmetry. In this state the $4f$ electronic degrees of freedom have to be included in the Fermi volume and hence the $4f$ electrons are fermions just like the valence electrons. Since this is a singlet state it has no net magnetic moment. The $4f$ electron's magnetic moment has been compensated by its anti-ferromagnetic coupling to the spins of the valence electrons. A question is: How to describe the two electron state at higher temperatures. At high temperatures the $4f$ electrons recover their localized magnetic moments and hence a singlet description is definitely not appropriate. Within quantum mechanical formalism, the low to high temperature transition in single ion Kondo systems can be modeled by a singlet to triplet transition wherein the high temperature state is the triplet state which has a magnetic moment [6, 7]. Such a treatment might explain the recovery of $4f$ magnetic moment at high temperatures. However within such a treatment the $4f$ electrons continue to remain as 'fermions' (indistinguishable particles) even at higher temperatures on account of their quantum mechanical spin triplet description thereby participating in the Fermi volume. Therefore, within this picture, the Fermi volume does not change with temperature.

In case of a Kondo lattice, Kondo phenomenon can be modeled as the coherent screening of the periodic 4*f* electrons by valence electrons creating a coherent Kondo state below the coherence temperature, $T_{coh}$ [8, 9]. The development of Kondo coherence leads to the loss of magnetic moments of the localized 4*f* electrons at low temperatures. The temperature dependent large-small Fermi surface transition in Kondo lattices is a long standing problem in the community. It is believed that the Fermi surface in Kondo systems should expand below $T_{coh}$ in order to incorporate 4*f* electronic degrees of freedom into it as a result of the Kondo coherence [10-12]. At high temperatures the Fermi surface remains small since the Kondo coherent state is not yet established and the 4*f* electrons are localized; hence they are excluded from the Fermi volume. Experimental evidence for such an expansion has been obtained in the de Haas van Alphen experiments performed on many Kondo lattices at very low temperatures [13]. These results have been successfully interpreted by considering the 4*f* electronic states as being a part of the Fermi volume and hence demonstrate the expanded character of the Fermi surface. The question is how the large-small Fermi surface transition should be modeled theoretically? There are few approaches based on dynamical mean field theory (DMFT) which have attempted to model the large-small Fermi surface transition in CeIrIn$_5$ [14, 15]. In this paper, we present an elegant conceptual picture to model this Fermi surface transition. We provide conditions under which electrons can/cannot be treated as fermions.

Within a quantum mechanical description of the Kondo phenomenon, the Fermi surface topology would undergo restructuring merely without any explicit expansion since the 4*f* electrons remain 'fermions' at all temperatures and hence their participation in the Fermi volume does not vary with temperature; a fact in contrast to experiments which have indeed demonstrated the large-small Fermi surface transition in Kondo lattices explicitly. Hence such a description of the Kondo phenomenon is inappropriate. Instead we claim that the high temperature state in a Kondo model is not a quantum mechanical state. For example, within the single ion Kondo model the high temperature state is not a spin triplet state but is a non-quantum state like $|\uparrow\rangle_f|\downarrow\rangle_v$ (or $|\downarrow\rangle_f|\uparrow\rangle_v$ equivalently since both states are degenerate in the absence of a magnetic field). This state does not have the exchange symmetry and therefore does not represent a quantum mechanical state describing indistinguishable particles. Instead, it is a state describing distinguishable particles wherein a particle exchange gives rise to a new state. We infer that 4$f$ electrons are not 'fermions' at high temperatures. Therefore we claim that the transition from the state $|\uparrow\rangle_f|\downarrow\rangle_v$ (or $|\downarrow\rangle_f|\uparrow\rangle_v$) at high temperatures to a state $|\uparrow\rangle_f|\downarrow\rangle_v - |\downarrow\rangle_f|\uparrow\rangle_v$ at low temperatures should be modeled 'semi-classically' but not quantum mechanically. Within this picture the large-small Fermi surface transition can be explained naturally. The state $|\uparrow\rangle_f|\downarrow\rangle_v - |\downarrow\rangle_f|\uparrow\rangle_v$ at low temperatures describes both 4$f$ and valence electrons as fermions participating in the Fermi volume that corresponds to a large Fermi surface while the state $|\uparrow\rangle_f|\downarrow\rangle_v$ (or $|\downarrow\rangle_f|\uparrow\rangle_v$) at high temperatures represents a localized 4$f$ electron which is excluded from the Fermi surface — the Fermi volume contains only the itinerant valence electron and hence the Fermi surface is small.

This conceptual picture may be easily visualized by comparing spatial extensions of wave functions for the localized 4$f$ and itinerant valence electrons. We argue that in order for an electronic system to be considered as a 'fermionic' system and treated within the Landau's Fermi liquid formalism, the spatial distribution of the amplitudes of the wave functions of all the constituent electrons must be identical to each other. Alternately one can state that for a fermionic system the spatial profile of the wave functions of all constituent electrons must be identical to each other. On the contrary when a 4$f$ electron is localized then it does not lead to such identical profile since the amplitude of 4$f$ electronic wave function peaks in the region of localization and becomes negligible far away from it whereas the amplitude of valence electronic wave function continues to be spatially uniform. Thus there is inconsistency between the spatial profiles of the wave functions of 4$f$ and valence electrons giving rise to 'distinguishability' between them. Hence the localized 4$f$ electron cannot be treated as a fermion. This happens in Kondo systems at high temperatures when the 4$f$ electron localizes. At low temperatures the spatial profile of the 4$f$ electronic wave function is identical to that of the valence electronic wave function due to quantum entanglement. Hence they both become indistinguishable and the 4$f$ electron is therefore a fermion. A remarkable consequence of this idea can be seen when tuning the ground state of Kondo lattices across QCP. When the Kondo ground state is formed in case of paramagnetic Kondo lattices the Fermi surface is large due to the occupancy of 4*f* electrons in the Fermi volume. The valence band contains both 4*f* and itinerant valence electrons and the Fermi liquid theory holds for the valence band. On the contrary when the 4*f* electron is completely localized in the magnetically ordered ground state of Kondo lattices then it cannot be treated as a fermion. Therefore the Fermi volume contains only the itinerant valence electrons excluding the 4*f* electrons (small Fermi surface) and the Fermi liquid theory still holds but only for the itinerant valence electrons which alone constitute the valence band (It is to be noted that the applicability of Fermi liquid theory is always in the context of valence band in solids since it is the valence electrons which influences the low energy excitations in solids. The 'localized' electrons do not contribute to the valence band and therefore cannot influence the low energy excitations in solids). When the system is tuned towards QCP then the degree of 4*f* electron delocalization is believed to increase until at QCP where we get *comparable* contribution from the 4*f* electron and the itinerant valence electrons in the valence band. Hence the low energy excitations of systems close to QCP are influenced jointly by the 4*f* and itinerant valence electrons. There are two possible theoretical scenarios currently believed for explaining the nature of 4*f* electrons in the vicinity of QCP. In one case it is predicted that the 4*f* magnetic moments are completely screened due to Kondo effect at QCP and the magnetic order results within the itinerant electron gas [16-18] whereas the other scenario predicts that the 4*f* electrons remain localized even at QCP and the magnetic order results from the localized electrons [19, 20]. All of these predictions are made by theoretical models within the quantum mechanical formalism. Either of these two scenarios have a certain amount of experimental support hence the debate over which among these two scenarios prevail is still ongoing. We provide a radically new theoretical picture in this regard. We argue that since the 4*f* electron transforms from a fermion in the paramagnetic ground state to a non-fermion (a distinguishable particle) in the magnetically ordered state through the QCP, the corresponding theoretical description for the transformation must be obtained in a semi-classical framework and not in a quantum mechanical framework. In this context we argue that the 4*f* electrons maintain their partial localized character even at the QCP which forbids their description as 'fermions'. Consequently the valence electronic system close to QCP, containing both the partially localized 4*f* (non-fermion) and itinerant valence electrons (fermion), cannot be modeled within the Landau's Fermi liquid formalism and hence NFL behavior emerges (For another explanation see Supplementary Information Section A).

Furthermore it is observed that the ground state close to QCP in many heavy fermion systems is superconducting in which the superconducting pairing mechanism is different from conventional phonon mediated. It is commonly believed that the unconventional superconducting pairing mechanism is electronic in origin. The question is why such an unconventional pairing mechanism is operative only close to QCP and not far away from it? We attempt to explain this in the following way. It is well known that every electronic system has a tendency to form a bosonic ground state in order to reduce its energy since bosons can condense into the same state at low temperatures whereas the fermions cannot do so due to Pauli's exclusion principle. It is due to this tendency that superconducting ground states are formed in case of certain metallic systems. However one needs a suitable mechanism to form superconducting Cooper pairs (bosons) out of electrons (fermions) in order to realize the superconducting ground state. We claim that the unconventional pairing mechanism is a consequence of the fact that the valence electrons for systems close to QCP are not fermionic in nature. Therefore the tendency to form a bosonic ground state in the valence electronic system is supported in such a case and superconductivity emerges. Far away from QCP the valence electrons are fermions and hence the unconventional pairing mechanism is not supported in such a case. Since Fermi liquid theory is not applicable close to QCP, quasi-particles can not be formed and therefore it is possible that the itinerant valence electrons find a glue to bind themselves together into a Cooper pair. Analogous to the case of phonon mediated superconductors where the glue is provided by the electron phonon coupling which causes an effective attractive interaction between two electrons giving rise to Cooper pairs [21, 22], the glue in case of heavy fermion superconductors seems to be generated by the temperature dependent attractive interaction between 4$f$ and itinerant valence electron – *Note that Kondo effect is an example of asymptotic freedom [23, 24] in condensed matter in which the strength of attractive coupling between 4f and valence electron is temperature dependent. The strength of this coupling increases with reducing temperature (see*

*Supplementary Information Section B). This coupling is a result of hybridization between 4f and valence electron. An effect of such temperature dependent coupling is clearly visible in the photoemission spectra as demonstrated in our earlier publication [25].* – effectively giving rise to an attractive interaction between two itinerant valence electrons binding them into a Cooper pair.

We anticipate that a similar reasoning holds while explaining the unconventional superconductivity observed in high $T_C$ Cuprate superconductors. Here too the metallic system above the superconducting transition temperature manifests anomalous behavior and therefore is called as a 'strange metal' [26, 27]. It is also believed that a QCP exists deep within the superconducting dome. Therefore it is quite likely that the unconventional pairing mechanism in Cuprate superconductor results from the fact that its valence electronic system is not fermionic in nature similar to the case of heavy fermion superconductors and hence unconventional superconductivity is supported in such a case. It is likely that the valence holes (electrons) in the hole (electron) doped Cuprate superconductors have an attractive temperature dependent coupling (resulting from the hybridization between the valence state and the localized magnetic moment) with the localized magnetic moment of the Cu ion (in $d^9$ electronic configuration) which gives rise to an effective attractive interaction between two valence holes (electrons) producing Cooper pairs causing high $T_C$ superconductivity — a similar picture as we proposed for explaining heavy fermion superconductivity.

### 3.    Conclusion

In summary, we argue that the high temperature state in the Kondo model cannot be treated within the quantum mechanical formalism. Instead we claim that the Kondo phenomenon should be modeled semi-classically. The semi-classical picture easily explains the experimentally observed large-small Fermi surface transition. The strength of our proposal comes from the fact that we highlight certain overlooked issues in quantum mechanics which led to inappropriate theoretical treatment of electronic states subjected to different degrees of localization. The key conceptual advance provided by this manuscript is the claim that not every electron qualifies to be called as a fermion and can be modeled within Landau's Fermi liquid formalism. Partially localized electrons alongside itinerant electrons together cannot be treated as a fermionic system and hence it displays NFL behavior in its low energy excitation spectrum. The non-fermionic nature of valence electronic system close to QCP supports the tendency of the electronic system to form a bosonic superconducting ground state and hence unconventional superconductivity is observed at QCP in heavy fermion systems. We project that a similar picture possibly holds for the explanation of unconventional superconductivity observed in high $T_C$ Cuprate superconductors as well. We argue that the Cooper pairing glue in case of heavy fermion or Cuprate superconductors results from the existence of an attractive temperature dependent coupling between valence states and the localized magnetic moment (resulting from the hybridization between the valence state and the localized magnetic moment) which can give rise to an effective attractive interaction between two valence electrons (or holes) forming Cooper pairs yielding unconventional superconductivity. We firmly believe that the qualitative/conceptual discussion presented in this manuscript would lay the foundation for the development of quantitative theoretical models addressing these issues.

# Supplementary Information

# A) Origin of Non-Fermi Liquid Behavior in heavy fermion systems

Landau's Fermi liquid theory claims that every many-electron system will possess fermionic quasiparticles as its low energy excitations no matter however large is the strength of electron correlations amongst them. Hence every electronic system will have a Fermi liquid ground state according to Landau. However, recent observations of non-Fermi liquid (NFL) behavior among many correlated electron systems have generated intense enquiry towards sorting out its origin.

In this manuscript, we argue that Landau overlooked a fundamental issue in quantum mechanics that an electronic system must possess identical spatial profile of wave function of every constituent electron in order to model the electronic system within the Fermi liquid theory. Such an identical profile ensures 'indistinguishability' amongst all the constituent electrons which is a necessary condition for them to be treated as 'fermions' obeying Fermi liquid theory. Such 'indistinguishability' can be realized in itinerant electron systems. These 'fermions' then give rise to the Fermi liquid ground state, regardless of the strength of correlations (e.g. as in paramagnetic heavy fermion systems). ***In the context of condensed matter systems, this means that the Fermi liquid theory is applicable for the itinerant 'valence' electrons but not for the localized electrons. In other words, itinerant 'valence' electrons are 'fermions'. Localized electrons are not 'fermions'.***

In the context of Kondo lattices, this idea gives rise to remarkable consequences: It is known that the ground state of a Kondo lattice possessing stable local magnetic moments is long range magnetically ordered. Here, only the itinerant valence electrons are fermions which alone constitute the valence band giving rise to Fermi liquid behavior. Localized 'magnetic' electrons are not fermions and do not participate in the valence band. On the other hand the Kondo screened ground state of a Kondo lattice is paramagnetic with all the localized 'magnetic' electrons becoming itinerant due to their quantum entanglement with the itinerant valence electrons. In this case, both the itinerant valence electrons and 'magnetic' electrons are fermions participating in the valence band. When we tune the ground state of a Kondo lattice towards the quantum critical point (QCP) from the magnetically ordered state (via variation of a control parameter like chemical/physical pressure, magnetic field etc.), we gradually introduce itinerancy in the localized 'magnetic' electrons. Consequently they start contributing to the valence band and affecting low energy physical properties of compounds close to QCP. However we claim that at QCP the Kondo screening process is not yet complete. Hence the 'magnetic' electrons possess partial localized character even at QCP which disqualifies them to be called as fermions. Consequently the valence band (containing both itinerant 'valence' electrons and partially localized 'magnetic' electrons) cannot be modeled within Fermi liquid theory and hence NFL behavior emerges at QCP.

# B) Temperature dependence of the Kondo coupling

The Hamiltonian for an impurity local magnetic moment $\vec{S}$ coupled to the conduction electron $\vec{s}$ within the Kondo model is:

$$H = \sum_{k\sigma} \varepsilon_k c^{\dagger}_{k\sigma} c_{k\sigma} - J\vec{S}\cdot\vec{s}$$

where

$$\vec{s} = \frac{1}{N} \sum_{k,k',\sigma,\sigma'} c^{\dagger}_{k\sigma} \vec{\sigma}_{\sigma,\sigma'} c_{k',\sigma'}$$ 

is the conduction electron spin at the impurity site ($\sigma$ are the Pauli matrices; $N$ is the number of sites)

$\varepsilon_k \rightarrow$ conduction electron dispersion

$c^{\dagger}_{k\sigma}, c_{k\sigma} \rightarrow$ creation and annihilation operator for an electron with wavevector $k$ and spin $\sigma$, respectively

$J \rightarrow$ Kondo exchange coupling constant ($J < 0$ for antiferromagnetic Kondo coupling between the local moment and the conduction electron spin)

{Note: Although the Kondo effect is known to be a many-body problem involving many electrons (~$10^{22}$ electrons/cm$^3$ - typical electron densities in metals) we attempt to illustrate the details about the Kondo coupling using an example of a two electron system (consisting of a localized magnetic electron and a conduction electron) under the justification that such details can be studied qualitatively using the two electron system. The effect of remaining electrons will only (at maximum) be to renormalize the value of the Kondo coupling $J$ and will not change the qualitative characteristics of $J$ like, e.g. its temperature dependence. **As a matter of fact the exchange interaction among electrons is by definition between two electrons.** Hence our use of such two electron system is well justified and is very helpful for illustrational simplicity.}

The prevalent view within the scientific community assumes $J$ to be constant with temperature. However, we argue that the value of $J$ increases with decreasing temperature corresponding to the increase in the Kondo coupling strength at lower temperatures. This can be understood qualitatively in the following way. It should be noted that $J$ is proportional to the exchange integral ($J_{ex}$) between the two electrons which is a function of their wavefunctions' spatial distribution as well as their mutual spatial separation (ref. http://en.wikipedia.org/wiki/Exchange_interaction). Specifically $J_{ex}$ is inversely proportional to the magnitude of their mutual spatial separation. As the temperature is lowered, due to the strong attractive interaction between the local moment and the conduction electron, there is a tendency for both of them to stay close to each other in space thus reducing their mutual spatial separation and thereby increasing the exchange integral $J_{ex}$ between the two electrons. Consequently the value of the Kondo exchange coupling $J$ increases too, at lower temperatures.

**Origin of an attractive interaction between a localized electron (undergoing finite hybridization with conduction electrons) and a conduction electron at low temperatures:**

The attractive nature of the exchange interaction $J$ can be qualitatively understood if one studies the physics of Hydrogen ($H_2$) molecule. In a $H_2$ molecule it is known that the bonding and anti-bonding states are formed due to the hybridization phenomenon between the two atomic H $1s$ orbitals. The bonding states involve anti-parallel orientation of both the spins while the anti-bonding states involve their parallel orientation. The bonding states correspond to the two electrons being close to each other in space while the opposite is true for the anti-bonding states (due to Pauli repulsion between both the electrons). The ground state is the bonding state which leads to the formation of the $H_2$ molecule while the anti-bonding state is an excited state which leads to the dissociation of the $H_2$ molecule. A similar situation exists between a localized electron and a conduction electron when we switch on the hybridization between them. Hybrid orbitals are formed as a consequence which share similar properties as those of the $H_2$ molecule. The resulting Kondo effect leads to the anti-ferromagnetic coupling between the two electrons at low temperatures leading to the formation of the Kondo singlet state at 0 K akin to antiparallel arrangement of the two electrons inside the bonding state of the $H_2$ molecule. Thus the localized electrons indeed contribute to the chemical bonding at low temperatures. In this state the electrons are very close in space. As we increase the temperature the anti-bonding states start getting populated thus reducing the occupancy of the bonding state.

*Thus the net effect of lowering temperature is the increase of the occupancy of the bonding state (with an antiferromagnetic interaction between the two electrons) causing a gradual reduction in the mutual spatial separation between both the electrons akin to the existence of an attractive interaction between both of them. Then following the formula for $J_{ex}$ from ref. http://en.wikipedia.org/wiki/Exchange_interaction, one can conclude that J increases with lowering of temperature due to such reduction. This J is responsible for generating the attractive force between both the electrons acting like glue for the formation of Cooper pairs in heavy fermion and Cuprate superconductors. Such is the microscopic origin of the attractive interaction between the localized electron and the conduction electrons at low temperatures.*

{Please note that the exchange coupling $J$ arises from the Coulomb interaction between both the electrons after applying quantum mechanical constraints on it and therefore does not represent a new fundamental force (unlike the electromagnetic force, gravitational force, strong nuclear force or weak nuclear force) in nature. Instead the exact effects of the Coulomb interaction among electrons were not fully taken into consideration so far and our aforementioned discussion merely fulfills this deficiency.}

We argue that *J*→0 as T→∞ and *J*→-∞ as T→0 (*J* < 0 always for antiferromagnetic Kondo coupling). *J*(T) is a monotonic function of temperature.

**Exchange Interaction and Fermionic behavior:**

Assume the two electron system to represent a localized magnetic 4*f* electron interacting with an itinerant valence electron by the Kondo interaction. At 0 K the system forms a stable non-magnetic Kondo singlet state denoted by: $|\uparrow\rangle_f|\downarrow\rangle_v - |\downarrow\rangle_f|\uparrow\rangle_v$. This state actually corresponds to *J*→-∞ wherein full exchange between both the electrons is realized. As a result we have full indistinguishability between them (their wave function being exchange invariant upto an overall phase factor). As the magnitude of *J* decreases with increasing temperature, the exchange interaction between both the electrons becomes weaker concomitantly. As a result we do not have full exchange between both the electrons instead only a *partial* exchange is realized between the two at non-zero finite temperatures. Consequently they are not fully indistinguishable between themselves but instead they should be considered as *partially* indistinguishable (what we mean by a *partial* exchange between the two electrons will be clear in the subsequent discussions). Thus we have a scenario in which the exchange interaction between the two electrons is temperature dependent. The *J*→0 state at T→∞ corresponds to the local moment state in which both the electrons become *completely* distinguishable. The question is how to express the two electron state at intermediate temperatures? We claim that an arbitrary temperature state could in general be written as: $|\uparrow\rangle_f|\downarrow\rangle_v - e^{i\theta}|\downarrow\rangle_f|\uparrow\rangle_v$, wherein the second term in the expression has acquired a phase factor θ different from that of the first term. Thus the weight of the second term which exchange interacts with the first term has reduced by a factor of cosθ as against the Kondo singlet state in which the second term exchange interacts with the first term completely. Such a situation amounts to a *partial* exchange between both the electrons giving rise to *partial* indistinguishability between them. Such a scenario cannot be captured within the quantum mechanical formalism and requires a semi-classical approach for its explanation. Note that θ should be expressed as θ = (π/2)*α where 0≤α≤1 and α depends on temperature via *J*(T). Thus α = α(*J*(T)) and α→0 as T→0 and α→1 as T→∞. α=0 corresponds to the Kondo singlet state while α=1 is the high temperature state where full local moment is obtained and which is devoid of antisymmetry (with respect to a particle exchange). Note that α should be monotonic function of temperature.

The fermionic behavior that we encounter in physics is a result of indistinguishability of the electrons. At any moment the Fermi surface (FS) volume of an electronic system will be proportional to the weight of the part of its wavefunction which obeys the antisymmetry (corresponding to indistinguishable electrons). Thus for the Kondo singlet state $|\uparrow\rangle_f|\downarrow\rangle_v - |\downarrow\rangle_f|\uparrow\rangle_v$, due to its full indistinguishability, we must count the *f*-electron wholly inside the FS. Thus the FS contains the valence electron (by default) and also the *f*-electron wholly {Please note that in the context of Kondo systems, the valence electrons are by default considered to be itinerant and participating in the FS volume. The open issues in Kondo systems concern the participation of the localized 4*f* electron within the FS volume with varying temperature and/or non-thermal 'control' parameter}. The FS volume corresponds to two electrons in that case. Now let us take an arbitrary temperature state i.e. $|\uparrow\rangle_f|\downarrow\rangle_v - e^{i\theta}|\downarrow\rangle_f|\uparrow\rangle_v$. We have already argued that this state does not correspond to full indistinguishability between both the electrons due to the absence of full antisymmetry (with respect to a particle exchange) for this state. However when we rewrite the above state as $(1-\cos\theta)|\uparrow\rangle_f|\downarrow\rangle_v + \cos\theta(|\uparrow\rangle_f|\downarrow\rangle_v - |\downarrow\rangle_f|\uparrow\rangle_v) - i\sin\theta|\downarrow\rangle_f|\uparrow\rangle_v$, one can see that a part of this state (i.e. the middle term) obeys the antisymmetry while the rest (i.e. the remaining terms) does not. It is here we claim that the state $|\uparrow\rangle_f|\downarrow\rangle_v - e^{i\theta}|\downarrow\rangle_f|\uparrow\rangle_v$ actually manifests *partial* indistinguishability (due to the existence of exchange interaction between a part of the 4*f* electron, i.e. cosθ fraction of the 4*f* electron, with the valence electron) between the two electrons. Only the part of the wavefunction which obeys the antisymmetry contributes to the FS volume while the rest does not contribute so and therefore it corresponds to the distinguishable electronic weight→ non-fermions {Please note that the distinguishability/indistinguishability of an electron is always defined in relation to some other electron. It is not obviously clear how this will affect the FS volume. Here we argue that typically in Kondo systems the valence electrons are considered as 'benchmark' for the electrons contributing to the FS volume. Therefore any other electron (e.g. 4*f* electron in Kondo systems) becoming indistinguishable with the valence electron must be counted in the FS volume and *vice versa*. The intermediate case of *partial* indistinguishability between them would naturally lead to a *partial* contribution of the 4*f* electron to the FS volume}. Thus in this case the total FS volume would contain the valence electron (by default) and cosθ fraction of one *f*-electron. Then the FS volume will be (1+cosθ). Thus one can see that with increasing temperature the FS volume reduces from 2 (at T = 0K) to 1 (at T = ∞K) following the expression (1+cosθ) = (1+cos((π/2)∗α)) which depends on the temperature through *J*(T). The exact functional form of *J*(T) is unknown and is an open and complex issue.

This was the illustration done for a simple two electron system highlighting the intricacies of the two electron exchange interaction as a function of temperature. Of course for real many-electron systems there might arise additional degrees of complexity and the exact FS evolution with temperature might need certain renormalization of the parameters presented here. However, the basic idea presented in this section (regarding the two electron exchange interaction) needs to be utilized in any realistic theory simulating the temperature dependent FS evolution in Kondo lattices.

**Non-fermi liquid (NFL) behavior at the quantum critical point (QCP):**

Instead of varying the temperature for tuning the value of *J*, one can also tune it by changing a non-thermal 'control' parameter (ρ) like physical/chemical pressure, applied magnetic field etc. Such a control parameter tuning has led to the discovery of QCP's in the heavy fermion systems which are a very active field of research currently. An interesting observation of NFL behavior close to QCP's has turned into a big open problem in contemporary condensed matter research.

We argue that our treatment about the state of the two electron system as described above is maintained even in the present case when we tune the ground state of the heavy fermion system by varying ρ. Along the ρ axis, (in case of the two electron system involving an interaction of the 4*f* electron with an effective field simulating the effect of other 4*f* sites of the Kondo lattice on the 4*f* electron → simulating the RKKY interaction (this is an approximation for the real situation in a Kondo lattice but it is quite helpful for highlighting our ideas over the variation of the two electron exchange interaction with ρ)) if ρ→∞ corresponds to $J(\rho)\to-\infty$ (Kondo singlet ground state, $|\uparrow\rangle_f|\downarrow\rangle_v - |\downarrow\rangle_f|\uparrow\rangle_v$) and ρ→0 corresponds to $J(\rho)\to 0$ (local magnetic moment ground state) then any ground state for a finite ρ will in general be represented by $|\uparrow\rangle_f|\downarrow\rangle_v - e^{i\theta}|\downarrow\rangle_f|\uparrow\rangle_v$ ,

where $\theta = (\pi/2)*\alpha$ ; $0\leq\alpha\leq 1$ and α depends on ρ via $J(\rho)$. Close to the critical value of ρ, i.e. $\rho_C$ where the QCP is believed to exist, the induced itinerancy into the *f*-electron (by increasing the magnitude of $J(\rho)$) is large enough for it to influence the physical properties of the systems although $\rho_C<\infty$. Thus the *f*-electron becomes a part of the valence band although it has not fully matured into a fermion yet. This leads the valence band to display NFL excitations.

In case of a Kondo lattice, the real situation is far more complex due to the RKKY interaction but the basic details about the two electron exchange interaction presented here must be utilized for a realistic simulation of the variation of the ground state of heavy fermion systems with ρ. This holds as well for the study of the variation of the FS volume with ρ, in which case we admit that a large amount of conflicting experimental evidence is available as far as the exact location of large-small FS transition with respect to the QCP is concerned. However our approach of treating the two electron exchange interaction within a semi-classical theory, we believe, would allow a large degree of flexibility for accommodating a wide variety of ground states within its predictions.